\documentclass[twocolumn]{aastex6}

\newcommand{\Rsun}{\ensuremath{{{\rm R}_\odot}}}
\newcommand{\Zsun}{\ensuremath{{Z_\odot}}}
\newcommand{\rev}[1]{{#1}}
\newcommand{\rrev}[1]{{#1}}

\usepackage{graphicx,lineno,svn-multi}
\usepackage{times}
\usepackage{mathptm}

\hypersetup{
  citecolor=black,     
}

\newcommand{\macro}[1]{\textcolor{red}{#1}}

\newcommand{\Msun}{\ensuremath{\mathrm{M}_\odot}}

\newcommand\OBSSTART{\macro{September 12}}
\newcommand\OBSEND{\macro{October 20}}
\newcommand{\OBSDAYS}{\macro{\ensuremath{\macro{16}~\mathrm{days}}}}
\newcommand\OBSEVENTFULLDATE{\macro{September 14, 2015 09:50:45 UTC}}

\newcommand{\CHIRPSTRAINPEAK}{\macro{\ensuremath{1.0 \times 10^{-21}}}}

\newcommand{\CHIRPDURATION}{\macro{\ensuremath{0.2}}}
\newcommand{\CHIRPFMIN}{\macro{\ensuremath{35}}}
\newcommand{\CHIRPFMAX}{\macro{\ensuremath{250}}}

\newcommand\OBSEVENTAPPROXCOMBINEDSNR{\macro{\ensuremath{24}}}

\newcommand{\APPROXPESKYNINTY}{\macro{\ensuremath{600~\mathrm{deg^2}}}}

\newcommand{\pergpcyr}{\ensuremath{\mathrm{Gpc}^{-3}\,\mathrm{yr}^{-1}}}
\newcommand{\ninetyinterval}[2]{\ensuremath{{#1}\text{--}{#2}}}

\newcommand{\purekklrateinterval}{\macro{\ensuremath{\ninetyinterval{2}{53}
    \, \pergpcyr}}}
\newcommand{\oneraterangetorulethemall}{\macro{\ensuremath{\ninetyinterval{2}{400} \, \pergpcyr}}}

\newcommand{\DISTANCECOMPACT}{\macro{\ensuremath{410_{-180}^{+160}}}} 
\newcommand{\MFINALSCOMPACT}{\macro{\ensuremath{62_{-4}^{+4}}}} 
\newcommand{\MONESCOMPACT}{\macro{\ensuremath{36_{-4}^{+5}}}} 
\newcommand{\MTWOSCOMPACT}{\macro{\ensuremath{29_{-4}^{+4}}}} 
\newcommand{\REDSHIFTCOMPACT}{\macro{\ensuremath{0.09_{-0.04}^{+0.03}}}} 
\newcommand{\SPINFINALCOMPACT}{\macro{\ensuremath{0.67_{-0.07}^{+0.05}}}} 
\newcommand{{\NCYCLES}}{{\macro{{\ensuremath{{10}}}}}} 
\newcommand{\MCobsCOMPACT}{\macro{\ensuremath{30_{-2}^{+2}}}} 
\newcommand{\MTOTobsCOMPACT}{\macro{\ensuremath{71_{-4}^{+5}}}} 
\newcommand{\MASSRATIOCOMPACT}{\macro{\ensuremath{0.82_{-0.21}^{+0.16}}}} 
\newcommand{\MCSCOMPACT}{\macro{\ensuremath{28_{-2}^{+2}}}} 
\newcommand{\DISTANCERANGE}{\macro{\ensuremath{230\text{--}570}}} 
\newcommand{\REDSHIFTRANGE}{\macro{\ensuremath{0.05\text{--}0.12}}} 

\renewcommand{\ninetyinterval}[2]{\ensuremath{{#1}-{#2}}}
\renewcommand{\macro}[1]{#1} 

\newcommand{\event}{GW150914}

\newcommand{\perGpcyr}{\ensuremath{\mathrm{Gpc}^{-3}\,\mathrm{yr}^{-1}}}

\begin{document}

\title{Astrophysical Implications of the Binary Black-Hole Merger
  GW150914} 
\author{%
B.~P.~Abbott,\altaffilmark{1}  
R.~Abbott,\altaffilmark{1}  
T.~D.~Abbott,\altaffilmark{2}  
M.~R.~Abernathy,\altaffilmark{1}  
F.~Acernese,\altaffilmark{3,4}
K.~Ackley,\altaffilmark{5}  
C.~Adams,\altaffilmark{6}  
T.~Adams,\altaffilmark{7}
P.~Addesso,\altaffilmark{3}  
R.~X.~Adhikari,\altaffilmark{1}
% ...
}  

\medskip
\affiliation {$^{\dag}$Deceased, May 2015. $^{\ddag}$Deceased, March 2015.
\\
{(LIGO Scientific Collaboration and Virgo Collaboration)}%
}%
\medskip

\altaffiltext {1}{LIGO, California Institute of Technology, Pasadena, CA 91125, USA }
\altaffiltext {2}{Louisiana State University, Baton Rouge, LA 70803, USA }
\altaffiltext {3}{Universit\`a di Salerno, Fisciano, I-84084 Salerno, Italy }
\altaffiltext {4}{INFN, Sezione di Napoli, Complesso Universitario di Monte S.Angelo, I-80126 Napoli, Italy }
\altaffiltext {5}{University of Florida, Gainesville, FL 32611, USA }
\altaffiltext {6}{LIGO Livingston Observatory, Livingston, LA 70754, USA }
\altaffiltext {7}{Laboratoire d'Annecy-le-Vieux de Physique des Particules (LAPP), Universit\'e Savoie Mont Blanc, CNRS/IN2P3, F-74941 Annecy-le-Vieux, France }
\altaffiltext {8}{Albert-Einstein-Institut, Max-Planck-Institut f\"ur Gravi\-ta\-tions\-physik, D-30167 Hannover, Germany }
\altaffiltext {9}{Nikhef, Science Park, 1098 XG Amsterdam, Netherlands }
\altaffiltext {10}{LIGO, Massachusetts Institute of Technology, Cambridge, MA 02139, USA }
\altaffiltext {11}{Instituto Nacional de Pesquisas Espaciais, 12227-010 S\~{a}o Jos\'{e} dos Campos, S\~{a}o Paulo, Brazil }
\altaffiltext {12}{INFN, Gran Sasso Science Institute, I-67100 L'Aquila, Italy }
\altaffiltext {13}{INFN, Sezione di Roma Tor Vergata, I-00133 Roma, Italy }
\altaffiltext {14}{Inter-University Centre for Astronomy and Astrophysics, Pune 411007, India }
\altaffiltext {15}{International Centre for Theoretical Sciences, Tata Institute of Fundamental Research, Bangalore 560012, India }
\altaffiltext {16}{University of Wisconsin-Milwaukee, Milwaukee, WI 53201, USA }
\altaffiltext {17}{Leibniz Universit\"at Hannover, D-30167 Hannover, Germany }
\altaffiltext {18}{Universit\`a di Pisa, I-56127 Pisa, Italy }
\altaffiltext {19}{INFN, Sezione di Pisa, I-56127 Pisa, Italy }
\altaffiltext {20}{Australian National University, Canberra, Australian Capital Territory 0200, Australia }
\altaffiltext {21}{The University of Mississippi, University, MS 38677, USA }
\altaffiltext {22}{California State University Fullerton, Fullerton, CA 92831, USA }
\altaffiltext {23}{LAL, Universit\'e Paris-Sud, CNRS/IN2P3, Universit\'e Paris-Saclay, 91400 Orsay, France }
\altaffiltext {24}{Chennai Mathematical Institute, Chennai 603103, India }
\altaffiltext {25}{Universit\`a di Roma Tor Vergata, I-00133 Roma, Italy }
\altaffiltext {26}{University of Southampton, Southampton SO17 1BJ, United Kingdom }
\altaffiltext {27}{Universit\"at Hamburg, D-22761 Hamburg, Germany }
\altaffiltext {28}{INFN, Sezione di Roma, I-00185 Roma, Italy }
\altaffiltext {29}{Albert-Einstein-Institut, Max-Planck-Institut f\"ur Gravitations\-physik, D-14476 Potsdam-Golm, Germany }
\altaffiltext {30}{APC, AstroParticule et Cosmologie, Universit\'e Paris Diderot, CNRS/IN2P3, CEA/Irfu, Observatoire de Paris, Sorbonne Paris Cit\'e, F-75205 Paris Cedex 13, France }
\altaffiltext {31}{Montana State University, Bozeman, MT 59717, USA }
\altaffiltext {32}{Universit\`a di Perugia, I-06123 Perugia, Italy }
\altaffiltext {33}{INFN, Sezione di Perugia, I-06123 Perugia, Italy }
\altaffiltext {34}{European Gravitational Observatory (EGO), I-56021 Cascina, Pisa, Italy }
\altaffiltext {35}{Syracuse University, Syracuse, NY 13244, USA }
\altaffiltext {36}{SUPA, University of Glasgow, Glasgow G12 8QQ, United Kingdom }
\altaffiltext {37}{LIGO Hanford Observatory, Richland, WA 99352, USA }
\altaffiltext {38}{Wigner RCP, RMKI, H-1121 Budapest, Konkoly Thege Mikl\'os \'ut 29-33, Hungary }
\altaffiltext {39}{Columbia University, New York, NY 10027, USA }
\altaffiltext {40}{Stanford University, Stanford, CA 94305, USA }
\altaffiltext {41}{Universit\`a di Padova, Dipartimento di Fisica e Astronomia, I-35131 Padova, Italy }
\altaffiltext {42}{INFN, Sezione di Padova, I-35131 Padova, Italy }
\altaffiltext {43}{CAMK-PAN, 00-716 Warsaw, Poland }
\altaffiltext {44}{Astronomical Observatory Warsaw University, 00-478 Warsaw, Poland }
\altaffiltext {45}{University of Birmingham, Birmingham B15 2TT, United Kingdom }
\altaffiltext {46}{Universit\`a degli Studi di Genova, I-16146 Genova, Italy }
\altaffiltext {47}{INFN, Sezione di Genova, I-16146 Genova, Italy }
\altaffiltext {48}{RRCAT, Indore MP 452013, India }
\altaffiltext {49}{Faculty of Physics, Lomonosov Moscow State University, Moscow 119991, Russia }
\altaffiltext {50}{SUPA, University of the West of Scotland, Paisley PA1 2BE, United Kingdom }
\altaffiltext {51}{University of Western Australia, Crawley, Western Australia 6009, Australia }
\altaffiltext {52}{Department of Astrophysics/IMAPP, Radboud University Nijmegen, P.O. Box 9010, 6500 GL Nijmegen, Netherlands }
\altaffiltext {53}{Artemis, Universit\'e C\^ote d'Azur, CNRS, Observatoire C\^ote d'Azur, CS 34229, Nice cedex 4, France }
\altaffiltext {54}{MTA E\"otv\"os University, ``Lendulet'' Astrophysics Research Group, Budapest 1117, Hungary }
\altaffiltext {55}{Institut de Physique de Rennes, CNRS, Universit\'e de Rennes 1, F-35042 Rennes, France }
\altaffiltext {56}{Washington State University, Pullman, WA 99164, USA }
\altaffiltext {57}{Universit\`a degli Studi di Urbino ``Carlo Bo,'' I-61029 Urbino, Italy }
\altaffiltext {58}{INFN, Sezione di Firenze, I-50019 Sesto Fiorentino, Firenze, Italy }
\altaffiltext {59}{University of Oregon, Eugene, OR 97403, USA }
\altaffiltext {60}{Laboratoire Kastler Brossel, UPMC-Sorbonne Universit\'es, CNRS, ENS-PSL Research University, Coll\`ege de France, F-75005 Paris, France }
\altaffiltext {61}{VU University Amsterdam, 1081 HV Amsterdam, Netherlands }
\altaffiltext {62}{University of Maryland, College Park, MD 20742, USA }
\altaffiltext {63}{Center for Relativistic Astrophysics and School of Physics, Georgia Institute of Technology, Atlanta, GA 30332, USA }
\altaffiltext {64}{Institut Lumi\`{e}re Mati\`{e}re, Universit\'{e} de Lyon, Universit\'{e} Claude Bernard Lyon 1, UMR CNRS 5306, 69622 Villeurbanne, France }
\altaffiltext {65}{Laboratoire des Mat\'eriaux Avanc\'es (LMA), IN2P3/CNRS, Universit\'e de Lyon, F-69622 Villeurbanne, Lyon, France }
\altaffiltext {66}{Universitat de les Illes Balears, IAC3---IEEC, E-07122 Palma de Mallorca, Spain }
\altaffiltext {67}{Universit\`a di Napoli ``Federico II,'' Complesso Universitario di Monte S.Angelo, I-80126 Napoli, Italy }
\altaffiltext {68}{NASA/Goddard Space Flight Center, Greenbelt, MD 20771, USA }
\altaffiltext {69}{Canadian Institute for Theoretical Astrophysics, University of Toronto, Toronto, Ontario M5S 3H8, Canada }
\altaffiltext {70}{Tsinghua University, Beijing 100084, China }
\altaffiltext {71}{Texas Tech University, Lubbock, TX 79409, USA }
\altaffiltext {72}{The Pennsylvania State University, University Park, PA 16802, USA }
\altaffiltext {73}{National Tsing Hua University, Hsinchu City, 30013 Taiwan, Republic of China }
\altaffiltext {74}{Charles Sturt University, Wagga Wagga, New South Wales 2678, Australia }
\altaffiltext {75}{University of Chicago, Chicago, IL 60637, USA }
\altaffiltext {76}{Caltech CaRT, Pasadena, CA 91125, USA }
\altaffiltext {77}{Korea Institute of Science and Technology Information, Daejeon 305-806, Korea }
\altaffiltext {78}{Carleton College, Northfield, MN 55057, USA }
\altaffiltext {79}{Universit\`a di Roma ``La Sapienza,'' I-00185 Roma, Italy }
\altaffiltext {80}{University of Brussels, Brussels 1050, Belgium }
\altaffiltext {81}{Sonoma State University, Rohnert Park, CA 94928, USA }
\altaffiltext {82}{Northwestern University, Evanston, IL 60208, USA }
\altaffiltext {83}{University of Minnesota, Minneapolis, MN 55455, USA }
\altaffiltext {84}{The University of Melbourne, Parkville, Victoria 3010, Australia }
\altaffiltext {85}{The University of Texas Rio Grande Valley, Brownsville, TX 78520, USA }
\altaffiltext {86}{The University of Sheffield, Sheffield S10 2TN, United Kingdom }
\altaffiltext {87}{University of Sannio at Benevento, I-82100 Benevento, Italy and INFN, Sezione di Napoli, I-80100 Napoli, Italy }
\altaffiltext {88}{Montclair State University, Montclair, NJ 07043, USA }
\altaffiltext {89}{Universit\`a di Trento, Dipartimento di Fisica, I-38123 Povo, Trento, Italy }
\altaffiltext {90}{INFN, Trento Institute for Fundamental Physics and Applications, I-38123 Povo, Trento, Italy }
\altaffiltext {91}{Cardiff University, Cardiff CF24 3AA, United Kingdom }
\altaffiltext {92}{National Astronomical Observatory of Japan, 2-21-1 Osawa, Mitaka, Tokyo 181-8588, Japan }
\altaffiltext {93}{School of Mathematics, University of Edinburgh, Edinburgh EH9 3FD, United Kingdom }
\altaffiltext {94}{Indian Institute of Technology, Gandhinagar Ahmedabad Gujarat 382424, India }
\altaffiltext {95}{Institute for Plasma Research, Bhat, Gandhinagar 382428, India }
\altaffiltext {96}{University of Szeged, D\'om t\'er 9, Szeged 6720, Hungary }
\altaffiltext {97}{Embry-Riddle Aeronautical University, Prescott, AZ 86301, USA }
\altaffiltext {98}{University of Michigan, Ann Arbor, MI 48109, USA }
\altaffiltext {99}{Tata Institute of Fundamental Research, Mumbai 400005, India }
\altaffiltext {100}{American University, Washington, D.C. 20016, USA }
\altaffiltext {101}{University of Massachusetts-Amherst, Amherst, MA 01003, USA }
\altaffiltext {102}{University of Adelaide, Adelaide, South Australia 5005, Australia }
\altaffiltext {103}{West Virginia University, Morgantown, WV 26506, USA }
\altaffiltext {104}{University of Bia{\l }ystok, 15-424 Bia{\l }ystok, Poland }
\altaffiltext {105}{SUPA, University of Strathclyde, Glasgow G1 1XQ, United Kingdom }
\altaffiltext {106}{IISER-TVM, CET Campus, Trivandrum Kerala 695016, India }
\altaffiltext {107}{Institute of Applied Physics, Nizhny Novgorod, 603950, Russia }
\altaffiltext {108}{Pusan National University, Busan 609-735, Korea }
\altaffiltext {109}{Hanyang University, Seoul 133-791, Korea }
\altaffiltext {110}{NCBJ, 05-400 \'Swierk-Otwock, Poland }
\altaffiltext {111}{IM-PAN, 00-956 Warsaw, Poland }
\altaffiltext {112}{Rochester Institute of Technology, Rochester, NY 14623, USA }
\altaffiltext {113}{Monash University, Victoria 3800, Australia }
\altaffiltext {114}{Seoul National University, Seoul 151-742, Korea }
\altaffiltext {115}{University of Alabama in Huntsville, Huntsville, AL 35899, USA }
\altaffiltext {116}{ESPCI, CNRS, F-75005 Paris, France }
\altaffiltext {117}{Universit\`a di Camerino, Dipartimento di Fisica, I-62032 Camerino, Italy }
\altaffiltext {118}{Southern University and A\&M College, Baton Rouge, LA 70813, USA }
\altaffiltext {119}{College of William and Mary, Williamsburg, VA 23187, USA }
\altaffiltext {120}{Instituto de F\'\i sica Te\'orica, University Estadual Paulista/ICTP South American Institute for Fundamental Research, S\~ao Paulo SP 01140-070, Brazil }
\altaffiltext {121}{University of Cambridge, Cambridge CB2 1TN, United Kingdom }
\altaffiltext {122}{IISER-Kolkata, Mohanpur, West Bengal 741252, India }
\altaffiltext {123}{Rutherford Appleton Laboratory, HSIC, Chilton, Didcot, Oxon OX11 0QX, United Kingdom }
\altaffiltext {124}{Whitman College, 345 Boyer Avenue, Walla Walla, WA 99362 USA }
\altaffiltext {125}{National Institute for Mathematical Sciences, Daejeon 305-390, Korea }
\altaffiltext {126}{Hobart and William Smith Colleges, Geneva, NY 14456, USA }
\altaffiltext {127}{Janusz Gil Institute of Astronomy, University of Zielona G\'ora, 65-265 Zielona G\'ora, Poland }
\altaffiltext {128}{Andrews University, Berrien Springs, MI 49104, USA }
\altaffiltext {129}{Universit\`a di Siena, I-53100 Siena, Italy }
\altaffiltext {130}{Trinity University, San Antonio, TX 78212, USA }
\altaffiltext {131}{University of Washington, Seattle, WA 98195, USA }
\altaffiltext {132}{Kenyon College, Gambier, OH 43022, USA }
\altaffiltext {133}{Abilene Christian University, Abilene, TX 79699, USA }

\begin{abstract}

The discovery of the gravitational-wave source \event\ with the
Advanced LIGO detectors provides the first observational evidence for
the existence of binary black-hole systems that inspiral and merge
within the age of the Universe. Such black-hole mergers have been
predicted in two main types of formation models, involving isolated
binaries in galactic fields or dynamical interactions in young and old
dense stellar environments. The measured masses robustly demonstrate
that relatively ``heavy'' black holes ($\gtrsim25\,$\,\Msun)
can form in nature.  This discovery implies relatively weak
massive-star winds and thus the formation of \event\ in an environment
with metallicity lower than $\simeq$\,1/2 of the solar value. The rate
of binary black-hole mergers inferred from the observation of
\event\ is consistent with the higher end of rate predictions
($\gtrsim 1$\,Gpc$^{-3}$\,yr$^{-1}$) from both types of formation
models. The low measured redshift ($z \simeq\,0.1$) of \event\ and the
low inferred metallicity of the stellar progenitor imply either binary black-hole
formation in a low-mass galaxy in the local Universe and a prompt
merger, or formation at high redshift with a time delay between
formation and merger of several\,Gyr. This discovery motivates further
studies of binary-black-hole formation astrophysics. It also has
implications for future detections and studies by Advanced LIGO and
Advanced Virgo, and gravitational-wave detectors in space.

\end{abstract}

\section{Introduction}

When in the 1970s the mass of the compact object in the X-ray binary
Cygnus X-1 was measured to exceed the maximum mass of a neutron star
(\citealt{1972Natur.235...37W,1972Natur.235..271B}), black holes (BHs)
turned from a theoretical concept into an observational reality.
Around the same time and over several years, evidence for supermassive
BHs in the centers of galaxies mounted
\citep[see][]{1995ARA&A..33..581K}. The formation of the stellar-mass
BHs found in X-ray binaries is associated with the core collapse (and
potential supernova explosion) of massive stars when they have
exhausted their nuclear fuel~\citep[e.g.][]{2003ApJ...591..288H}. The
origin of supermassive BHs is less clear. They may have small seeds,
that originate from ``heavy'' stellar-mass BHs (more massive than
$\simeq$\,25\,\Msun) or large seeds formed from intermediate-mass BHs
formed in the earliest generations of massive stars or directly from
large clouds \citep[see][]{2010A&ARv..18..279V}.

The gravitational-wave (GW) signal \event\ detected on
\OBSEVENTFULLDATE\ by the Advanced LIGO (aLIGO)
detectors~\citep{GW150914-DETECTION} has been shown to originate from the
coalescence of a binary BH (BBH) with masses of
\MONESCOMPACT\,\Msun\ and \MTWOSCOMPACT\,\Msun\ (in the source frame,
\rev{see \S\,\ref{properties}}). This GW discovery provides the first
robust confirmation of several theoretical predictions: (i) that
``heavy'' BHs exist, (ii) that BBHs form in nature, and (iii) that
BBHs merge within the age of the Universe at a detectable rate.

The inspiral and merger of binaries with BHs or neutron stars (NSs)
have been discussed as the primary source for ground-based GW
interferometers for many decades
\citep[e.g.,][]{thorne.k:1987,1989CQGra...6.1761S}. The existence of
GWs was established with radio observations of the orbital decay of
the first binary pulsar, PSR B1913+16
\citep{1975ApJ...195L..51H,1982ApJ...253..908T}. Even before the
binary pulsar discovery, \citet{1973NInfo..27...70T} described the
evolution of isolated massive binaries (i.e. those not influenced
dynamically by surrounding stars) and predicted the formation of
binary compact objects that merge, including BBHs. Some of the first
\emph{population} studies of massive stellar binaries and their
evolution even predicted that BBH mergers could dominate detection
rates for ground-based GW interferometric
detectors~\citep{1997MNRAS.288..245L}. Furthermore,
\citet{1993Natur.364..423S} recognized that dense star clusters
provide another possible way of forming merging BBHs: BHs in dense
star clusters quickly become the most massive objects, sink towards
the cluster core, subsequently form pairs through dynamical
interactions, and are most commonly ejected in binary configurations
with inspiral times shorter than the age of the Universe. For the most
recent review articles on the formation of binary compact objects in
galactic fields and dense stellar systems,
see~\cite{2014LRR....17....3P} and~\cite{2013LRR....16....4B},
respectively.

In this paper we discuss \event\ in the context of astrophysical
predictions in the literature and we identify the most robust
constraints on BBH formation models. In \rev{\S\,~\ref{properties} we
  report the properties of \event, in} \S\,~\ref{masses} and
  \S\,~\ref{redshift} we discuss the implications of the measured BH
  masses and distance to the source. In \S\,~\ref{e_spin} and
  \S\,~\ref{rates} we examine conclusions that can be drawn from the
  GW constraints on the orbital eccentricity, BH spins, and BBH merger
  rates. In \S\,~\ref{future} we discuss prospects for future
  detections and the types of astrophysical studies we would need to
  further advance our understanding of BBH formation. In
  \S\,~\ref{conclusions} we summarize our key conclusions.

\section{\rev{The properties of \event}}\label{properties}

\rev{\event\ was discovered first through a low-latency search for
  gravitational-wave transients, and later in subsequent match-filter
  analyses of \OBSDAYS\ of coincident data collected by the two aLIGO
  detectors between \OBSSTART\ to \OBSEND\ \citep{GW150914-DETECTION}. The
  signal matches the waveform expected for the inspiral, merger, and
  ringdown from a compact binary. In \CHIRPDURATION\ s it swept in
  frequency from \CHIRPFMIN\ to \CHIRPFMAX\ Hz, reaching a peak GW
  strain of \CHIRPSTRAINPEAK\ with a signal-to-noise ratio of
  \OBSEVENTAPPROXCOMBINEDSNR\ \citep{GW150914-BURST,GW150914-CBC}. }

\rev{Consideration of these basic signal properties of frequency and
  frequency derivative indicate \rrev{that} the source is a BH merger. Coherent
  Bayesian analyses~\citep{PE} using advanced waveforms
  \citep{Taracchini:2013rva,Puerrer:2014fza,Hannam:2013oca,Khan:2015jqa}
  allow us to measure several of the source physical parameters (all
  quoted at 90\% credible level). In the detector frame, the chirp
  mass\footnote{The chirp mass is $\mathcal{M} = (m_1 m_2)^{3/5}/(m_1
    + m_2)^{1/5}$, where $m_1$ and $m_2$ are the component masses.}
  is \MCobsCOMPACT\,\Msun\, and the total mass is
  \MTOTobsCOMPACT\,\Msun; the mass ratio is \MASSRATIOCOMPACT\,\Msun\,
  and the luminosity distance is determined to be
  \DISTANCECOMPACT\,Mpc (redshift \REDSHIFTCOMPACT). The two BH masses
  in the source frame then are \MONESCOMPACT\,\Msun\ and
  \MTWOSCOMPACT\,\Msun, and the chirp mass in the source frame is
  \MCSCOMPACT\,\Msun. The source-frame mass and spin of the final BH
  are \MFINALSCOMPACT\,\Msun\ and \SPINFINALCOMPACT\ and the source is
  localized to a sky area of \APPROXPESKYNINTY\ \citep[see also][]{GW150914-PARAMESTIM,
    GW150914-EMFOLLOW}. The signal does not show deviations from the
  expectations of general relativity, as discussed in detail
  in~\citet{GW150914-TESTOFGR}.}
  
\rev{ Assuming that the source-frame BBH merger rate is constant
  within the volume in which \event\ could have been detected (found
  to extend out to redshift of $\simeq0.5$) and that \event\ is
  representative of the underlying BBH population, the BBH merger rate
  is inferred to be \purekklrateinterval\, in the comoving frame at
  the 90\% credible level~\citep{Kim:2002uw,GW150914-RATES}. The match-filter
  searches of these 16\,days of coincident data also revealed a number
  of sub-threshold triggers with associated probabilities of them
  being astrophysical or noise in nature~\citep{GW150914-CBC}. If we
  account for the probability of these sub-threshold triggers and we
  consider a wide range of models for the underlying BBH mass
  distribution, the estimated BBH merger rates extend to the
  range~\oneraterangetorulethemall\,~\citep{2015PhRvD..91b3005F,GW150914-RATES}.}

\section{Black-Hole Masses in Merging Binaries}\label{masses}

\subsection{Brief Review of Measured BH Masses}

Prior to the discovery of \event, our knowledge of stellar-mass
BH masses has come from the study of X-ray binaries (XRBs) where a
compact object accretes matter from a stellar companion
\citep[e.g.,][]{McClintock:2006}.
Dynamical compact-object mass measurements in these binaries rely on
measurements of the system's orbital period, the amplitude of the
stellar-companion's radial velocity curve, and quantitative
constraints on the binary inclination and the companion mass
\citep[e.g.,][]{2014SSRv..183..223C}. When the mass of the compact
object is found to exceed 3\,\Msun, which is the conservative upper
limit for stable \rev{NS~\citep{1974PhRvL..32..324R,1996ApJ...470L..61K}}, then the XRB is
considered to host a black hole. At present 22 BH XRBs have confirmed
dynamical mass measurements, 19 of these systems lie in our
Galaxy. For the majority of the systems, the measured BH masses are
5--10\,\Msun, while some have masses of 10--20\,\Msun\footnote{For
  probability distribution functions of measured BH masses,
  see~\citet[][]{2011ApJ...741..103F,2010ApJ...725.1918O}.}. Black
holes have been claimed to be measured dynamically in two other
extragalactic systems, IC10 X-1~\citep[][$M_{\rm BH}$ =
  21--35\,\Msun]{2007ApJ...669L..21P,2008ApJ...678L..17S} and NGC300
X-1~\citep[][$M_{\rm BH}$ = 12--24\,\Msun]{2010MNRAS.403L..41C}. On
the basis of these observations, \citet{2011ApJ...730..140B} argue
that these two systems are likely immediate progenitors of BBH
systems. However, recent work casts these BH masses in doubt: it now
appears more likely that the measured velocities are due to
stellar-wind features instead of the BH companion's orbital
motion~\citep[and references therein]{2015MNRAS.452L..31L}, and
therefore we do not consider the claimed BH masses in these systems as
reliable.

All these observed BH systems are found in low stellar density
galactic fields. Based on multi-wavelength electromagnetic studies of
X-ray point sources, BH XRBs have also been claimed to exist in
globular clusters~\citep[][and
  references therein]{2007Natur.445..183M,2013ApJ...777...69C};
however, dynamical mass measurements for these systems have not been
possible, and hence reliable BH mass constraints are not available.

  \emph{Both BHs of the
  \event\ coalescence are more massive than the BHs in known XRBs with
  reliably measured mass: this GW discovery provides the most robust
  evidence for the existence of ``heavy'' ($\gtrsim25$\,\Msun)
  stellar-mass BHs.} In what follows we review our current
understanding of BH and BBH formation both in isolation and in dense
environments, and we examine the implications of the high \event\ BH
masses.

\subsection{Predicted Masses for Single BHs}

\begin{figure*}
  \includegraphics[width=\columnwidth]{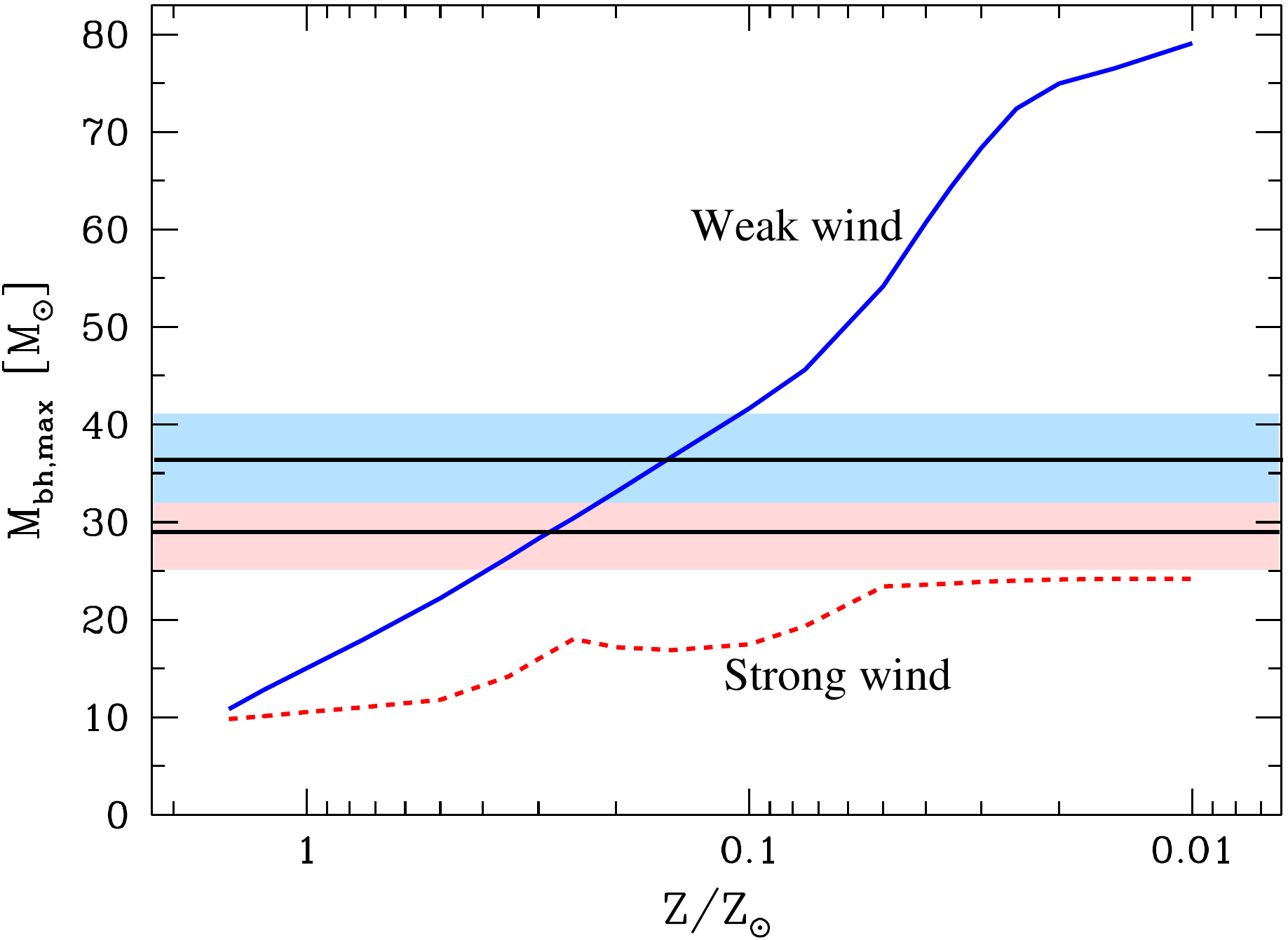}
  \hspace*{0.3cm}
  \includegraphics[width=0.98\columnwidth]{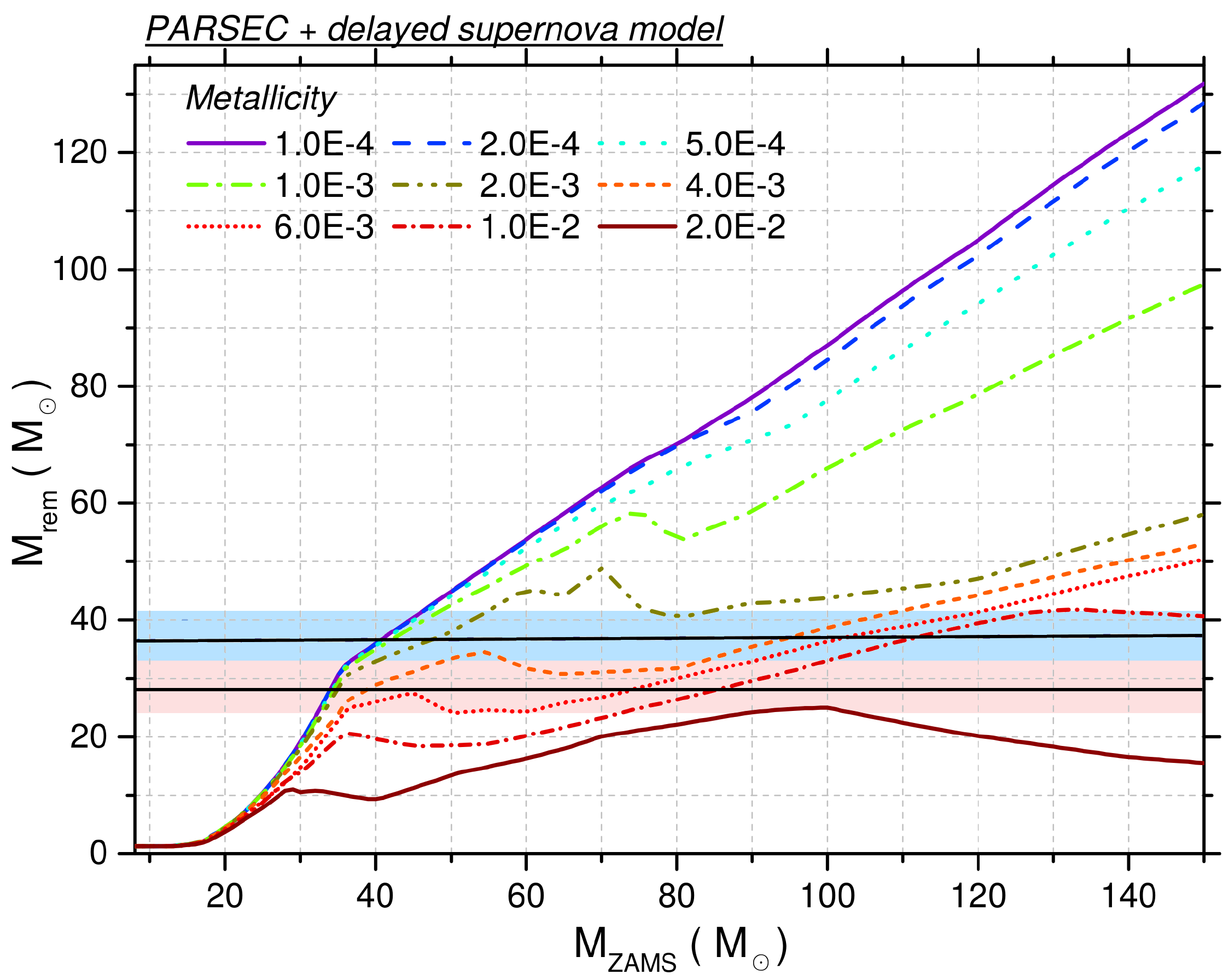}
  \caption{\label{fig:mass-met} Left: dependence of maximum BH mass on
    metallicity $Z$, with $Z_\odot = 0.02$ for the old (strong) and
    new (weak) massive star winds \citep[Figure 3
      from][]{2010ApJ...714.1217B}. Right: compact-remnant mass as a
    function of zero-age main-sequence (ZAMS; i.e., initial)
    progenitor mass for a set of different (absolute) metallicity
    values \citep[Figure 6 from][]{2015MNRAS.451.4086S}. The masses of
    \event\ are indicated by the horizontal bands.}
\end{figure*}

Black holes are expected to form at the end of the nuclear lifetimes
of massive stars. The stellar core collapses to form a proto-NS and
the occurrence and strength of a supernova (SN) explosion determines
how much material is ejected, and whether a BH is
formed. \citet{2001ApJ...554..548F} distinguish BH formation through
partial or full fallback of the initially exploding envelope, or
through the complete collapse of the BH progenitor due to a core
collapse that is not powerful enough to drive an
explosion. \citet{2012ApJ...749...91F} find that the transition from
NS to BH formation occurs at initial progenitor masses of
$\simeq$\,18--20\,\Msun~ and the transition from fallback to complete
BH collapse takes place at initial progenitor masses of
$\simeq$\,40\,\Msun. Other studies
\citep[e.g.,][]{2012ApJ...757...69U} find that either the SN is
successful and a NS is formed, or the whole star collapses to a BH;
there is a range of progenitor masses (15--40\,\Msun\,for solar
metallicity) where either a NS or a BH could form.

This relatively simple picture of BH formation from single stars is
greatly affected by several key factors: the strength of massive-star
winds and their dependence on the star's metallicity \citep[$Z$,
  e.g.,][]{1992A&A...264..105M}, stellar rotation
\citep[e.g.,][]{2009A&A...497..243D}, and the microphysics of stellar
evolution. These factors affect the relationship between the initial
progenitor mass and the stellar (core) mass at the time of collapse,
and thus the mass of the resulting BH.

Winds are understood to be radiation-driven. Their strengths depend on
stellar  properties, but cannot be derived from first
principles; instead they are empirically derived and calibrated based
on observations \citep[for a review, see][]{2014ARA&A..52..487S}. Over
the last decades, wind strengths for different evolutionary stages
have been significantly revised, mainly downwards leading to more
massive progenitors at core collapse~\citep[for a review
  see][]{2008NewAR..52..419V}. In general, stars at lower metallicities
exhibit weaker winds, since the lower metal content reduces opacity,
enables easier radiation transport, and reduces radiation momentum
transfer and hence mass loss from the stellar surface. The functional
dependence on metallicity is also empirically constrained by studying
massive stars in environments of different metallicities. However, the
range in metallicities probed by observations is much smaller than the
range where massive stars are formed over cosmic history, and hence
extrapolations to metallicities orders of magnitude smaller than solar
\Zsun\ (i.e., $Z=0.02$) are adopted.  Although we have no way of
validating such extrapolations, here we consider the published
low-metallicity models.

\citet{2003ApJ...591..288H} and \citet{2009MNRAS.395L..71M} were among the first
to examine how compact object formation depends on progenitor masses,
stellar winds, and metallicity, albeit in a rather qualitative
framework. Quantitatively, \citet{2010ApJ...714.1217B} and later
\citet{2013MNRAS.429.2298M,2015MNRAS.451.4086S} showed that adopting
the latest wind prescriptions~\citep{2008NewAR..52..419V}
significantly increases the stellar mass at core collapse and thus the
maximum BH mass that can form from single stars, although the
exact relation between initial mass and final BH mass depends on the
details of the wind prescription
(see Figure~\ref{fig:mass-met}).

Stellar rotation can lead to angular momentum transport and extra
mixing in the stellar interiors. In extreme cases, the evolution of the
star can be significantly altered, avoiding expansion of the star to
giant dimensions \citep{1987A&A...178..159M}. It has been proposed
that rapid rotation, especially at low metallicities, where winds and
associated angular momentum losses are weaker, or in close binaries,
where tides may replenish the angular momentum, may play a significant
role in the formation of more massive
BHs~\citep{2009A&A...497..243D,2016arXiv160100007M,2016arXiv160103718M}.
Nevertheless, there are no calculations that find BHs more massive
than 30\,\Msun\, unless the metallicity is lower than \Zsun. 

Stellar properties at core collapse and the ensuing compact-remnant
masses have also been shown to depend, albeit much more weakly, on the
treatment of microphysics in stellar structure and evolution codes,
especially on assumptions regarding convective overshooting and
resultant
mixing~\citep{2015MNRAS.447.3115J}. Finally,~\citet{2012ApJ...749...91F}
and~\citet{2015MNRAS.451.4086S} investigate how basic properties of
the supernova explosion might affect remnant masses at different
metallicities. They show that remnant masses in excess of
$\simeq$\,12\,\Msun\, at \Zsun\ ($\simeq$\,30\,\Msun~ at 1/100 \Zsun)
are formed through complete collapse of their
progenitors. Therefore, the masses of BHs in ``heavy'' BBH
  mergers only carry information about the evolution leading up to the
  collapse and not about the supernova mechanism.

The measured masses of the merging BHs in \event\ show that
stellar-mass BHs as massive as 32\,\Msun\, (the lower limit on the
more massive BH at 90\% credible level) can form in nature. {\em Given
  our current understanding of BH formation from massive stars, using
  the latest stellar wind, rotation, and metallicity models, we
  conclude that the \event\ BBH most likely formed in a
  low-metallicity environment: below $\simeq$1/2\,\Zsun\, and possibly
  below
  $\simeq$\,1/4\,\Zsun~\citep{2010ApJ...714.1217B,2013MNRAS.429.2298M,2015MNRAS.451.4086S}. }

It is, in principle, possible that ``heavy'' BHs are formed
  through indirect paths that do not require low metallicity, but we
  consider this very unlikely. For example, the formation of ``heavy''
  BHs through the dynamical mergers of lower-mass BHs
  with massive stars in young clusters has been considered.  However,
  these models adopt the optimistic assumption that in such mergers,
  even for grazing collisions, all of the mass is retained, leading to
  significant BH mass
  growth~\citep{2014ApJ...794....7M,2014MNRAS.441.3703Z}. Stellar
  collisions in dense stellar environments
  (see~\citealt{1999A&A...348..117P}) could potentially produce stars
  massive enough to form ``heavy'' BHs, but these objects are also
  subject to strong winds and intense mass loss unless they are stars
  of low metallicity ~\citep{2009A&A...497..255G}. Finally, formation
  of ``heavy'' BHs from the mergers of lower-mass BHs in clusters is
  unlikely because most dynamically formed merging BBHs are ejected
  from the host cluster before merger \citep[][see their
    Figure~2]{2015PhRvL.115e1101R}.

\subsection{BBH Masses from Isolated Binary Systems}\label{mass_binaries}

The fact that the majority of massive stars are members of binary
systems \rev{with a roughly flat mass-ratio distribution}
\citep{2007ApJ...670..747K,2012Sci...337..444S,2014ApJS..213...34K}
provides the opportunity for BBH formation in isolated binary
systems. In that case, the masses of BHs depend not only on the
initial mass of the star and metallicity, but also on any binary
interactions. The development of binary population models focused on
the formation of double compact objects goes back
to~\citet{1983SvA....27..334K} and~\citet{1987ApJ...321..780D}, but
the first population models to account for BBH formation appeared a
decade later starting with~\citet{1993MNRAS.260..675T}. Several groups
have explored different aspects of BBH formation from isolated
binaries at varying levels of detail~(many reviewed
by~\citealt{2007PhR...442...75K,2009NewAR..53...27V,2014LRR....17....3P}).
Models find that BBH formation typically progresses through the
following steps: (i) stable mass transfer between two massive stars,
although potentially non-conservative (i.e., with mass and
angular momentum losses from the binary), (ii) the first core collapse
and BH formation event, (iii) a second mass transfer phase that is
dynamically unstable leading to inspiral in a common envelope (in
which the first BH potentially grows slightly in mass;
\citealt{2005ApJ...632.1035O}), (iv) the second core-collapse event
leading to BBH formation, and (v) inspiral due to GW emission and
merger. \citet{2012ApJ...759...52D} found that the vast majority of BBH
mergers follow this evolutionary path: 99\% at solar metallicity and
90\% at 0.1\,\Zsun. Alternative formation pathways, avoiding mass
transfer and common envelope, may be possible if massive stars remain
rapidly rotating, stay chemically homogeneous through their lifetimes,
remain compact, and do not become giant stars
\rrev{\citep[][]{2009A&A...497..243D,2016arXiv160100007M,2016arXiv160103718M}.}

Most studies indicate that model predictions, in particular merger
rates, but also probability distributions of BBH properties,
are affected by a considerable number of physical factors and
associated parameters, albeit at different levels of sensitivity: (i)
initial binary properties (masses, mass ratios, and orbital periods),
(ii) stellar evolution models including metallicity-dependent
wind-driven mass loss, (iii) mass and associated angular momentum
transfer between binary components and loss from the systems, (iv)
treatment of tidal evolution, (v) treatment of common-envelope
evolution, and (vi) BH natal kicks. The significance of (v) and
  (vi) has been discussed recently for the
StarTrack~\citep{2008ApJS..174..223B} models by
\citet{2012ApJ...759...52D,2015arXiv151004615B}. Recently,
\citet{2015arXiv150603573D} concluded that the current uncertainties
in initial binary properties (i) do not dramatically change the
rates. The other factors, i.e., (ii) -- (vi), have been
consistently identified as important, not just for rate predictions,
but also for predictions of BH mass spectra in merging BBHs.

As we have discussed, the \event\ masses favor the newer, weaker
stellar winds and metallicities below \Zsun. Quantitative models for
BH and BBH formation considering such conditions have appeared only in
the past five years, starting with~\citet{2010ApJ...715L.138B}, and
in numerous follow-up
studies~\citep{2012ApJ...759...52D,2013ApJ...779...72D,2015ApJ...806..263D,2015MNRAS.451.4086S,2015arXiv151004615B}. \cite{2013ApJ...779...72D}
fold in cosmological effects, accounting for redshift evolution of the
formation rate and metallicity (down to $Z =10^{-4}$). With the
extension to such low metallicities, the strong dependence on the
common-envelope treatment found earlier \citep{2012ApJ...759...52D} is
weakened in the case of formation of BHs more massive than
20\,\Msun. In fact, it is striking that, once full metallicity
evolution is included, BBH systems that merge within the age of the
Universe and have total masses as high as $\sim$\,100\,\Msun\, are
rather generically formed regardless of other model assumptions;
still, predicted detectable samples seem to be dominated by less
massive BBH systems~\citep{2015ApJ...806..263D,2014ApJ...789..120B}.

On the extreme low-metallicity end, it has been proposed that BBH
formation is also possible in the case of stellar binaries at zero
metallicity (Population III, PopIII, stars;
see~\citealt{2004ApJ...608L..45B,2014MNRAS.442.2963K}). The
predictions from these studies are even more uncertain, since we have
no observational constraints on the properties of first-generation
stellar binaries (e.g., mass function, mass ratios, orbital
separations). However, if one assumes that the properties of
PopIII massive binaries are not very different from binary
populations in the local Universe (admittedly a considerable
extrapolation), then recently predicted BBH total masses agree
astonishingly well with \event\ and can have sufficiently long merger
times to occur in the nearby Universe~\citep{2014MNRAS.442.2963K}.
This is in contrast to the predicted mass properties of low (as
opposed to zero) metallicity populations, which show broader
distributions~\citep{2015arXiv151004615B}.

\emph{We conclude that predictions from a broad range of models for
  BBH formation from isolated binaries are consistent with the
  \event\ masses provided newer, weaker massive-star winds and
  extrapolations to metallicities of 1/2\,\Zsun\ or lower are
  adopted.} More calculations of massive binary evolution with updated
wind prescriptions and taking cosmological evolution into account are
needed to fully exploit the new information that would be provided by
additional GW detections.

\subsection{BBH Masses from Dense Stellar Environments} \label{mass_clusters}

Over the last few decades our understanding of the evolution of BHs in
dense stellar clusters has evolved considerably. Based on early
analyses~\citep{1993Natur.364..421K,1993Natur.364..423S} BHs form in
clusters from massive stars and quickly \rev{mass segregate to the
  center through dynamical friction \rrev{(on a timescale shorter than the
  overall relaxation time by a factor that is the ratio of the mass of
  the typical BH mass to the average background star mass)}.} In these
high-density conditions, BHs dynamically interact, forming binaries,
and often are ejected from the cluster. \rev{Such dynamical
  interactions preferentially keep the heaviest objects in binaries
  and eject the lightest, producing heavier binaries and driving mass
  ratios closer to unity~\citep{1975MNRAS.173..729H}.}
\citet{2000ApJ...528L..17P} presented the first significant $N$-body
simulation of equal-mass BHs in a dense cluster, and they found that
the ejected BBH systems are sufficiently eccentric that they will
merge within the age of the Universe at a rate important for
LIGO/Virgo observations. Since then, studies of varying levels of
detail have examined BBH formation in clusters and have identified the
importance of three-body interactions for hardening binaries to the
point they can merge in a Hubble time, pointing out that these
interactions are also responsible for dynamical
ejections~\citep{2004ApJ...616..221G,2006ApJ...640..156G,2006ApJ...648..411K,2010MNRAS.402..371B,2014MNRAS.440.2714B}
as well as in galactic centers
\citep{MillerLauburg:2008,2009MNRAS.395.2127O,2012PhRvD..85l3005K,2013ApJ...777..103T}. Gravitational-wave
kicks~(\citealt{2015PhRvD..92b4022Z} and references therein) can also
eject post-merger, single BHs from their host clusters. Throughout
these studies BHs are assumed to be of a single fixed mass (typically
10\,\Msun). That means that, although their results are relevant for
our understanding of the physics of stellar dynamics on BBH formation
and evolution and the expected merger rates (section~\ref{rates}),
they cannot be used to determine the expected masses of BBH mergers
formed in dynamical environments.

\citet{2006ApJ...637..937O,OLeary:2007} and
\citep{2008ApJ...676.1162S} presented the first BBH population predictions
from dense clusters with a BH mass spectrum. Their treatment of the
effects of stellar dynamics was based on simple cross sections and a
static density background. Nevertheless, their results generically
produced BBH mergers in the local Universe with BH masses of several
tens of solar masses.

The first simulations to account in detail for both binary evolution
and stellar dynamics with a BH mass spectrum and with realistic
numbers of particles were by
\citet{2010MNRAS.407.1946D,2011MNRAS.416..133D} and by
\citet{2013ApJ...763L..15M,2015ApJ...800....9M}. \cite{2015ApJ...800....9M,PhysRevLett.116.029901}
further accounted for a population of globular clusters with varying
cluster properties (mass, density, and metallicity). Examination of
these results indicates, very much like the models of isolated binary
evolution, that clusters of lower metallicity produce BBH mergers of
higher masses, with chirp masses in excess of 10\,\Msun and up to
25--30\,\Msun\, (the chirp mass of \event\ is \MCSCOMPACT\,\Msun). We
note that none of these studies incorporate the newer, weaker winds
leading to more massive BHs adopted by some of the models for isolated
binaries (section~\ref{mass_binaries}). Such a modification applied to
clusters will unavoidably increase the BBH masses from clusters even
further.  BBH populations are also predicted to form in young, open
clusters~\citep{2014ApJ...781...81G,2014MNRAS.441.3703Z} with
``heavy'' masses. In this case, BBHs are formed mostly through
dynamical exchanges in three-body encounters of single BHs with
binaries containing one or two BHs.

\emph{We conclude that BBH formation in dense star clusters is
  consistent with \event, provided that the clusters have typical
  metallicities lower than \Zsun\, in order to form sufficiently massive
  merging BBHs. Most of these mergers occur outside the clusters
  following dynamical BBH ejection.} Although, under optimistic
assumptions, formation of ``heavy'' BHs at solar metallicity has been
discussed through stellar or BH-star mergers in young clusters, such
paths require chains of dynamical interactions for eventual BBH
formation.  Cluster models with more detailed treatment of binary
evolution, dynamics, updated stellar winds, and exploration of cluster
properties are needed to determine the robustness of
the predicted BBH mass spectra.

\section{Binary Black-Hole Mergers in the Nearby Universe}\label{redshift}

\begin{figure*}
  \includegraphics[height=\textwidth,angle=-90]{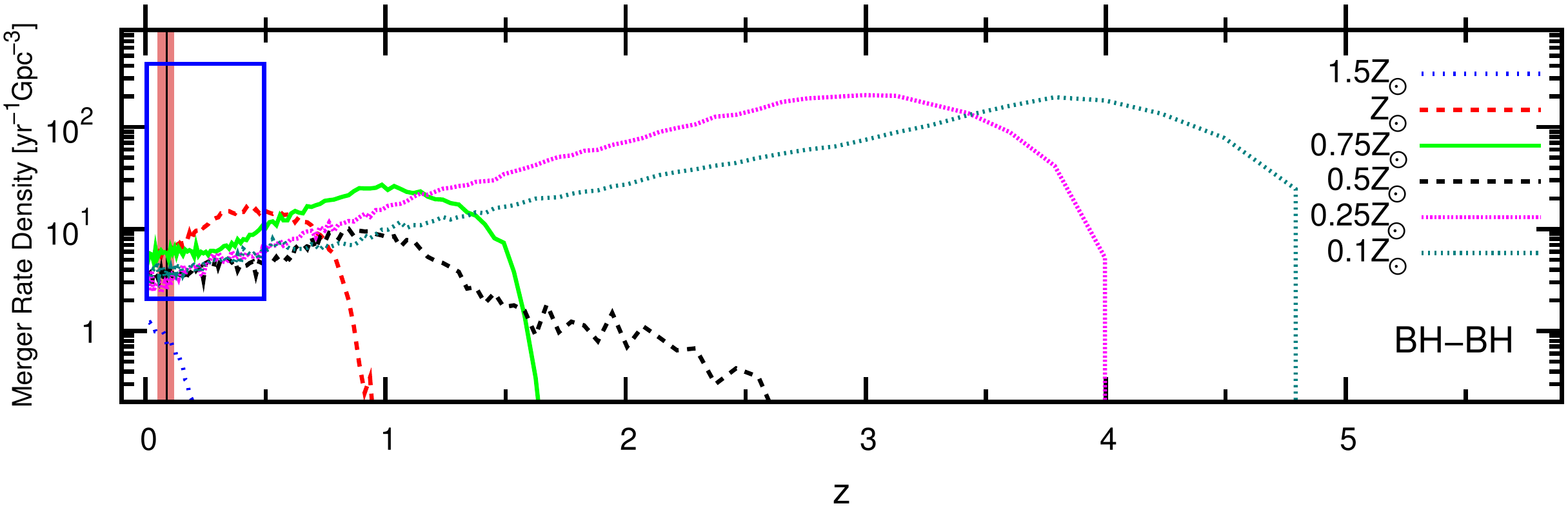}
  \caption{\label{fig:rate-Z} Predictions of BBH merger rate \rev{in
      the comoving frame} (\perGpcyr) from isolated binary evolution
    as a function of redshift for different metallicity values
    (adopted from Figure~4 in \citealt{2013ApJ...779...72D}). \rev{At a given redshift, the
      total merger rate is the sum over metallicity.} The redshift range
    of \event\ is indicated by the vertical band; the range of the BBH
    rate estimates and the \rev{redshift out to which a system like 
      \event\ could have been detected in this observing period are} indicated
    by an open blue rectangular box.}
\end{figure*}

\begin{figure}
  \includegraphics[width=\columnwidth]{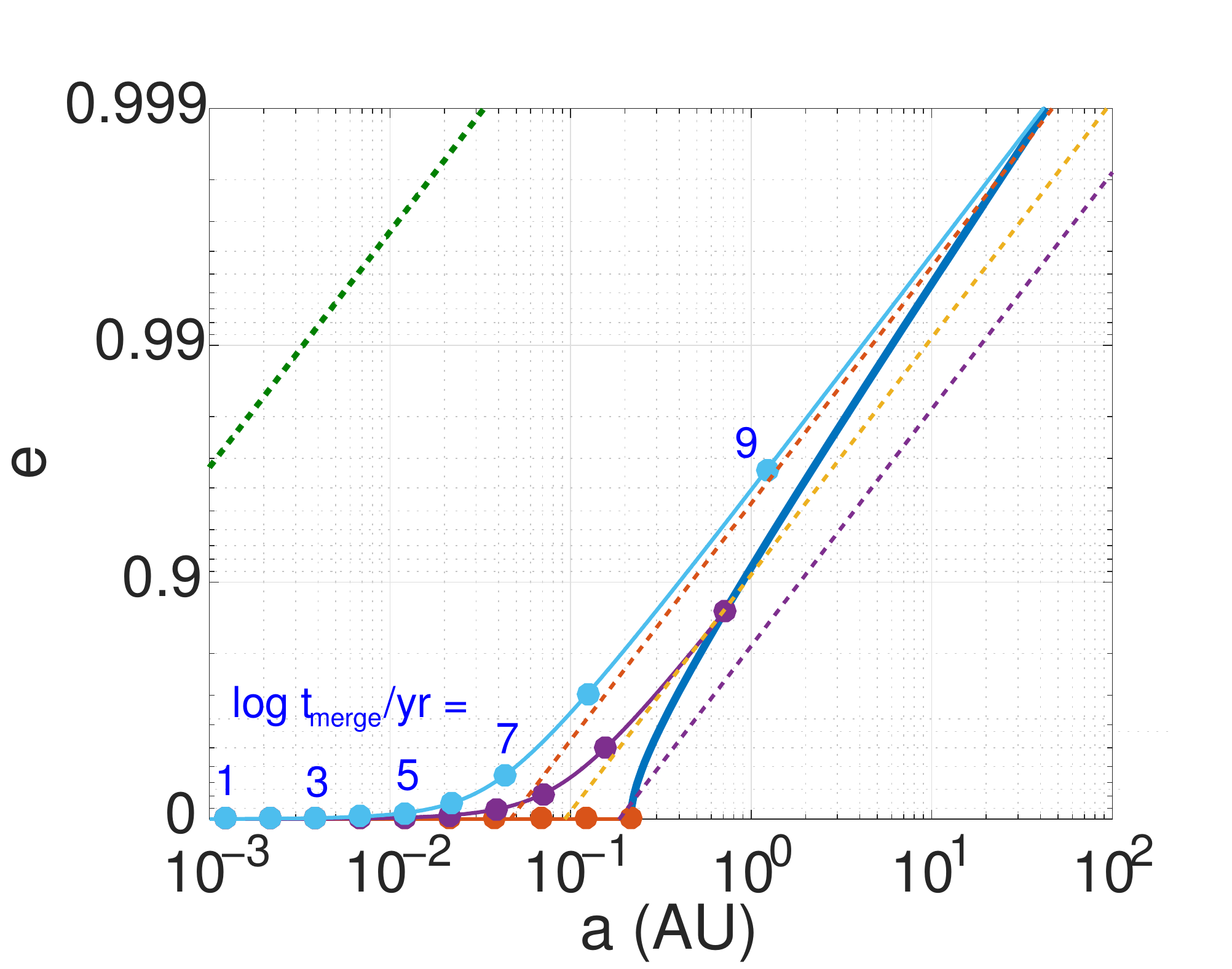}
  \caption{\label{fig:a-e} Allowed initial BBH semimajor axis and
    eccentricity in order to merge within 10\,Gyr (left of the thick
    solid blue line) \rev{for a BBH with the 
      \event\ masses}. The thin solid lines with circles represent the
    evolutionary trajectories of individual example systems, starting
    at the edge of the allowed range (the circles give the time to
    merger of $\log t$/yr = 1, 2, 3, 4 ... 10, from left to right).
    The dashed lines denote periastron separations of 10, 20, and 40
    \,\Rsun\ (left to right: orange, yellow, purple). The green dotted
    line shows the trajectory of a binary that has a remaining
    eccentricity of 0.1 at a GW frequency of 10 Hz.}
\end{figure}

Apart from the BH masses of the binary system, another important
measurement of \event\ is its luminosity distance in the range of
\DISTANCERANGE\,Mpc (at 90\% credible level) which corresponds
to a redshift of \REDSHIFTRANGE\, and an age of the Universe of
$\simeq 12.2-13.1$\,Gyr at the time of the merger (using \citealt{2015arXiv150201589P}).  The
specific implications of this measurement vary, depending on whether
\event\ originated from isolated binary evolution or from dynamical
interactions in a dense stellar environment.

\emph{In the case of dynamical origin, mergers of such
  ``heavy'' BBHs in the local Universe fit
  comfortably.} Models of dynamical BBH
formation~\citep{2000ApJ...528L..17P,MillerLauburg:2008,2009MNRAS.395.2127O,2011MNRAS.416..133D,2012PhRvD..85l3005K,2013ApJ...777..103T,2014MNRAS.441.3703Z,2015PhRvL.115e1101R,PhysRevLett.116.029901,2015ApJ...800....9M}
show that stellar and BH interactions take about $\sim$\,1\,Gyr to
form BBHs which have a wide range of delay times between formation and
merger for BBHs from old and young clusters.
  
{\em In the case of a BBH merger from an isolated binary at low
  metallicity, \rev{there is a continuum of possibilities in between two extremes: the BBH
      progenitor of \event\ could have formed in the local universe
      with a short merger delay time, or it could have formed at
      higher redshift with a long merger delay time. We cannot
      distinguish between these two extremes with the observation of this
      single event.}}

Short merger times are typically favored. One of the most recent
  isolated binary model predictions \citep[][]{2013ApJ...779...72D},
  shows preference for merger times of $\simeq$\,10--300\,Myr. However,
  low-metallicity star formation is rare in nearby galaxies. The age
and metallicity distribution of a large sample of nearby galaxies from
the Sloan Digital Sky Survey (SDSS) with median redshift of 0.13 shows
that very few galaxies have low metallicities, all of which are
low-mass ($<$\,$10^9$\,\Msun) and have relatively young stellar
populations~\citep[$<$\,1\,Gyr,][]{2005MNRAS.362...41G}.
In any case, the well measured mass-metallicity relation
\citep{2004ApJ...613..898T} implies that BBH formation paths with
merger delay times below $\sim$\,1\,Gyr require that the source
originated in nearby ($z$\,$<$\,0.2), low-mass, young galaxies. 

Alternatively, the BBH system may have formed much earlier (e.g.,
$z$\,$\gtrsim$\,2), when low-metallicity star formation was more
common \citep[see][]{2014ARA&A..52..415M}, but then it must have taken
much longer to merge ($\sim$\,5--10\,Gyr). Such long merger delay
times are often disfavored significantly compared to short delays
  \cite[by factors of 10--100; see, e.g.][]{2013ApJ...779...72D}.

To present a more quantitative discussion, we consider the study by
\citet{2013ApJ...779...72D} in more detail. They discuss BBH formation
from isolated binaries, accounting for the dependence on star
formation, galaxy-mass, and metallicity evolution from the local
Universe to cosmological redshifts and find that most local BBH
mergers originate from star formation in the first few Gyr of the
Universe and with long merger delay times \citep[see
  also][]{2015ApJ...806..263D}. Figure~\ref{fig:rate-Z} \citep[adopted
  from Figure\,4 of][]{2013ApJ...779...72D} indicates that the BBH
merger rate of binaries with metallicities of 1/2, 1/4 and 1/10
Z$_\odot$ increases with redshift and peaks at redshifts of 1, 3 and
4, respectively, i.e., at distances much larger than the measured
\event\ luminosity distance. The local ($z\sim$\,0.1) BBH merger rates
at such low metallicities are suppressed by factors of $\sim$\,10--100
compared to higher redshifts, but they are still comparable (within a
factor of about 2) to the high-metallicity local merger rate densities
that produce lower-mass BBHs.

To further study the potential progenitors of \event\ and their
expected merger time, we plot in Figure~\ref{fig:a-e} the allowed
parameter range for the initial (right after BBH formation) semimajor
axis ($a$) and eccentricity ($e$) of the BBH orbit that produces a
merger within 10 Gyr, using the point-mass approximation of
\citet{Peters:1964}. Binaries with long delay times originate close to
the thick solid line. Evolutionary trajectories show that systems
become circular long before merger, even for high initial $e$, unless
they form with extremely short merger times or extremely high $e$ (see
section~\ref{e_spin}).  For initially circular orbits, $a$ needs to be
smaller than 0.215\,AU or 46\,\Rsun\, \rev{for the binary to merge
  within $\sim$10\,Gyr}.  BBHs that form from two existing BHs in
clusters can form anywhere in the allowed parameter range. In the case
of isolated binaries, the \rev{separation before the formation of the second BH
  needs to be wide enough to accommodate the progenitor star. The BBH then forms with a similar separation (or similar periastron distance, if there is mass loss in the supernova or if BHs
  receive natal kicks), unless the BH kick is large and fine-tuned in its direction to drastically change the orbital separation. Since these progenitor stars} have
radii of at least several \,\Rsun\ ($\gtrsim 10$\,\Rsun\ for
chemically homogeneous evolution), we estimate that the periastron
distance needs to be larger than $\sim$10--20 \,\Rsun\ as indicated in
Figure~\ref{fig:a-e}.

{\em We conclude that, based on published model results, ``heavy'' BBH
  mergers from low-metallicity environments in the local Universe are
  not particularly surprising, regardless of whether their origin is
  dynamical or from isolated binaries. The rate of ``heavy'' BBH
    mergers may very well increase with redshift either due to the
    increase in low-Z star-formation rates or due to higher rates at
    shorter merger times, at least for redshifts of up to $\simeq$\,1. These redshifts are within the horizon distance of aLIGO/Advanced-Virgo (AdV) design sensitivity, expected to be reached by
    $\sim$\,2020~\citep[][and see \S\,\ref{future}]{Aasi:2013wya}. }

\section{Binary Eccentricity and Black-Hole Spins}\label{e_spin}

There is no evidence for eccentricity in the orbital dynamics of
\event, but \rev{ eccentricities of $\lesssim$0.1 would
  not be detectable for this event~\citep{GW150914-PARAMESTIM}}. In any case, from
Figure~\ref{fig:a-e} it is clear that any eccentricity would have
dissipated by the time the binary entered the detectors' sensitive
frequency band. Indeed, in this Figure we plot the evolution of a
system that would retain an eccentricity of 0.1 at 10 Hz, but this
evolution only takes 1.25 days from $e = 0.999$ to merger!
\citet{2011A&A...527A..70K} and \citet{2002ApJ...572..407B} show that
for their field BBH models, the expected eccentricities would be
undetectable. Only formation in a dynamical environment at short
semimajor axis and an extremely high eccentricity could produce a
detectable eccentricity \citep[e.g.][]{2009MNRAS.395.2127O} for a
small fraction of BBHs. A small fraction of BBHs may form through
triple stars in globular clusters and potentially maintain significant
eccentricities until the merger~\citep{Samsin2014,
  2015arXiv150905080A}.

Parameter-estimation analysis of \event\ \citep{GW150914-PARAMESTIM} with gravitational
waveforms that account for spin effects (including precession)
constrains the dimensionless spin magnitude of the primary BH to
$\lesssim$\,0.7 (at 90\% credible level); the spin of the secondary BH
is not significantly constrained. The dimensionless spin components
aligned (or anti-aligned) with the orbital angular momentum axis are
likely to be small, whereas the spin components in the orbital plane
are poorly constrained. The tentative implication is that, if spin
magnitudes are indeed large, then the spin-orbit misalignment is
likely to be high too; if the spin magnitudes are small, then the
tilts remain unconstrained.
  
These BH spin magnitude constraints derived from GW observations are
comparable in strength to what we typically obtain from X-ray
analyses~\citep[for reviews,
  see][]{2014SSRv..183..295M,2015PhR...548....1M}. These BH spin
estimates in XRBs have been made from analysis of the X-ray spectra of
accretion disks, based either on the influence of a spin-dependent
radius of the disk inner edge on the continuum of the spectra or of
the effect of the BH spin on the shape of emission lines. Black-hole
spins are typically found to be high for systems with high-mass
donors. In general, the cores of massive stars are expected to rotate
rapidly and thus may lead to rapid BH spin at formation, unless there
is efficient angular momentum coupling between the core and the
(expanding) envelope \citep[e.g.][]{2005A&A...443..581H}. The ability
to constrain the BH spins in \event\ reveals a new approach to
understanding the spin distribution of BHs that is independent of XRB
measurements.
Measuring BH spins in a variety of BH binaries has the potential of
revealing the origin of BH spins, at formation and through subsequent
accretion evolution in
binaries~\citep[e.g.][]{2008ApJ...682..474B,2010Natur.468...77V,2012ApJ...747..111W,2015ApJ...800...17F,2015arXiv151204897A}.

 For BBH formation from isolated binaries, BH spin alignment is
 expected if the spin of the BH is aligned with the spin of its
 progenitor star and thus with the binary. Even if BH kicks are
 relatively large ($>$100\,km\,s$^{-1}$), it is found that BBH spin
 tilts are rather constrained to typical values below about
 $\simeq$\,20\,deg~\citep{2000ApJ...541..319K}.  For BBH formation in
 dense environments there is no reasoning suggesting that spins would
 be correlated in any way through BH interactions and thus significant
 misalignment would be more likely.  Thus, if we would know that the
 BH spins in \event\ were aligned with the orbital angular momentum,
 then their magnitude would already be constrained by this GW
 measurement.  \rev{Conversely, spin precession effects significantly
   modify the relative orientation of the two BH spins between their
   formation and
   merger~\citep{Schnittman:2004,2015PhRvL.115n1102G,Kesden:2015},
   particularly when both spin magnitudes are large.  While initially
   random spins remain random at coalescence, spin precession effects
   can distort the relative likelihood of some misalignment
   angles. Therefore the misalignment angles measured for
   \event\ cannot be directly identified as the birth BH spin
   misalignments.}

\emph{We conclude that the non-detection of eccentricity for
  \event\ is not a surprise regardless of the BBH formation
  mechanism. \rev{Since the spin magnitude is not expected to change during the X-ray binary phase, the upper limit on the primary BH spin indicates
    that it was not formed with extremal spin}. At
  present, the evidence for relatively small magnitudes of the BH spin
  components aligned with the orbital angular momentum does not
  provide constraints on the formation mechanism. The non-aligned
  components and hence spin-orbit tilts are essentially
  unconstrained. With additional BBH detections, a clear preference
  for mostly aligned spins would favor formation from isolated
  binaries and small natal BH kicks. On the other hand, \rev{a predominance of large misalignments could favor formation through dynamical processes instead.} As the sample
  grows, spin measurements will prove critical for distinguishing
  formation channels \rev{and their relative contributions to the
    merger rate}. }

\begin{figure*}
  \includegraphics[width=1.07\columnwidth]{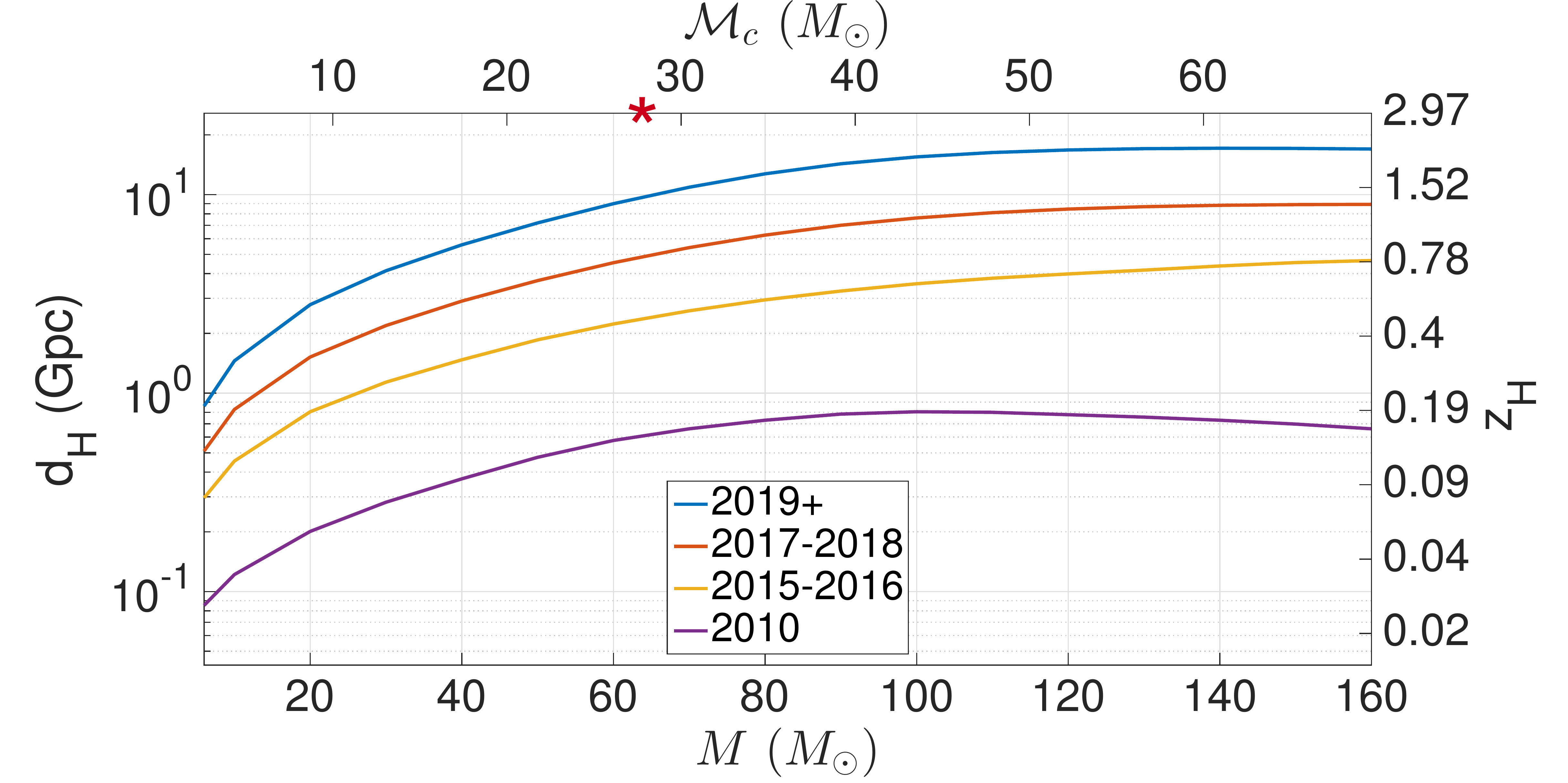}
  \includegraphics[width=1.09\columnwidth]{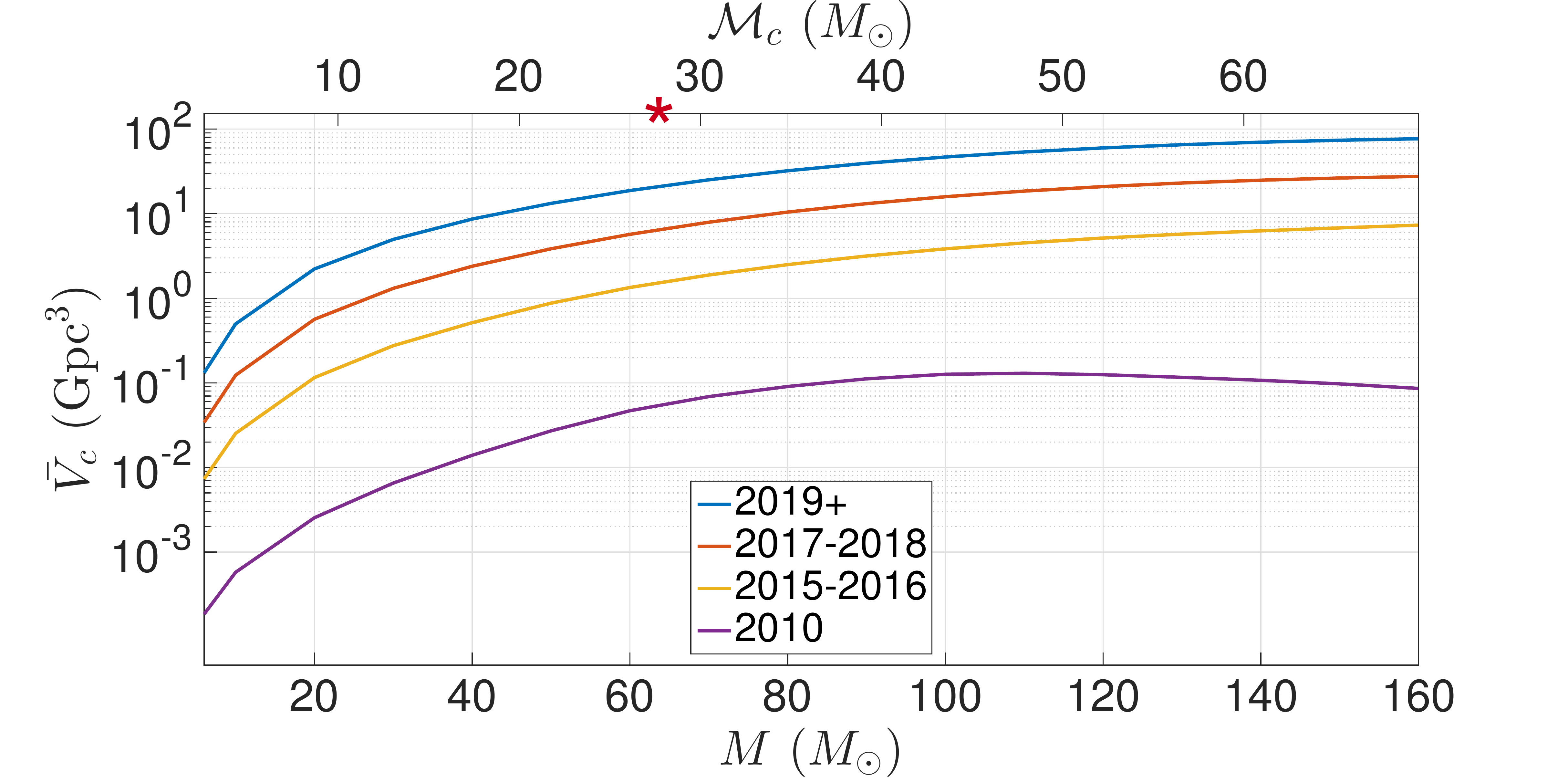}
  \caption{\label{fig:DL_M} Left: Horizon distance (left axis) and
    horizon redshift (right axis) as a function of total mass (bottom
    axis) and chirp mass (top axis), for equal mass\rev{, non-spinning}
  BBH mergers. The (expected) increase in detector sensitivity with
  time is shown by the different lines \rev{and the chirp mass of
    \event\ is indicated with a red star}. Right: the same, but now for
  detection-weighted sensitive comoving volume, defined to yield the
  expected number of detections if multiplied with a merger rate per
  unit volume. For details see
  Appendix.}
\end{figure*}

\section{Binary Black-Hole Merger Rates}\label{rates}

The upper limits on the merger rates from initial LIGO/Virgo
observations were not stringent enough to exclude even the most
optimistic theoretical predictions~\citep{Aasi:2012rja}. In contrast,
GW150914 provides the first interesting GW rate constraints on
astrophysical models. \rev{As discussed in \S\,~\ref{properties}, the rate at which such BBH
mergers occur in the low-redshift Universe ($z \lesssim$0.5) is inferred in the range of $2-400$~\perGpcyr~\citep{GW150914-RATES}.}

Over the years, some studies have discussed models of isolated binary
populations, which result in completely aborting the formation of BBH
systems that merge within the age of the Universe,
e.g.,~\citet{Nelemans:2001,2002ApJ...572..407B} and most
recently~\citet{2014A&A...564A.134M}. In all these models, the lack of
BBH mergers can be traced back to one or more of the following model
assumptions: strong (old) wind models; no metallicity dependence of
wind strengths; no orbital evolution due to tides; high BH natal
kicks. All these assumptions effectively widen the orbits of massive
binaries and prevent, not the formation of BBH systems in general, but
more specifically the formation of BBH systems that merge within the
age of the Universe. Dynamical formation of BBHs is aborted if BHs receive natal kicks larger than
the local escape speed \citep[e.g., $\gtrsim$50 km/s for typical
  globular clusters, see][]{2002ApJ...568L..23G} such that the BHs
escape before they can interact.

{\em The existence of GW150914 shows that BBH mergers occur in nature,
  and therefore models which don't predict their existence within a
  Hubble time through any formation channel are excluded~(e.g.,
  certain models in
  \citealt{Nelemans:2001,2002ApJ...572..407B,2007ApJ...662..504B,2014A&A...564A.134M}).
  For both isolated binary evolution and dynamical formation, the
  implication of BBH existence is that BH kicks cannot always be high ($>$\,100\,km\,s$^{-1}$), in order to avoid disrupting or
  widening the orbits too much, or ejecting the BHs from clusters
  before they can interact.  In the case of isolated binaries, BBH
  existence also implies that massive star winds cannot be strong, and
  in the absence of high rotation, survival through common-envelope
  evolution in massive binaries must be possible.}

Rate predictions for binary mergers and associated LIGO/Virgo
detection expectations were summarized in \citet{ratesdoc}, and for
BBH mergers a range of $0.1-300$ \perGpcyr was reported. More recent
studies, not included in \citet{ratesdoc}, for isolated binary
evolution give very similar predictions: $0 -
100$~\perGpcyr\ by~\citep{2016arXiv160100007M}, $0.5 -
220$~\perGpcyr\ by~\citet{2015ApJ...806..263D},
$0-1000$~\perGpcyr\ by~\citet{2014A&A...564A.134M}. Recent studies of
globular cluster dynamics also report comparable rates
\citep{2015PhRvL.115e1101R,PhysRevLett.116.029901,2010MNRAS.407.1946D,2011MNRAS.416..133D}.\emph{We
  conclude that the \event\ rate constraints are broadly consistent
  with most of the BBH rate predictions, and only the lowest predicted
  rates ($\lesssim1$~\perGpcyr) can be excluded.}

\section{The Path Forward for Future Studies}\label{future}

In the coming years the aLIGO and AdV detectors will be upgraded to
a higher sensitivity, as shown in Figure~\ref{fig:DL_M}: on the left we
plot the maximum luminosity distance ($D_L$) and redshift ($z$), and
on the right a measure for the surveyed volume ($\overline{V}_c$) for
the initial LIGO/Virgo detectors, the current aLIGO and future
expectations (see the Appendix for the details). We can anticipate that
the BBH detection sample will increase by at least a factor of
$\sim$\,10 as observing runs become more sensitive and of longer
duration. With these new detections, it will become possible to go
beyond the mostly qualitative inferences discussed here, and
quantitatively constrain the properties of double-compact-objects
(DCOs) and their formation models.

In general, quantitatively constraining the model can be done either
by deriving a parametrized description of the underlying model
\citep[e.g.][]{Mandel:2009pc,2013PhRvD..88h4061O} or by comparing
specific population models to the data
\citep[e.g.][]{BulikBelczynski:2003,Mandel:2009nx}.

For the latter, detailed information about the models and properties
of the predicted populations are needed, e.g., masses and rate
densities as a function of redshift. Given the large number of model
parameters, it is challenging to obtain a statistically appropriate
sampling of the parameter space to the level required to address
degeneracies; no existing study has provided a sufficiently complete
data set.  However, such analyses will eventually allow us to constrain
massive-star winds and rotation, the common-envelope binary evolution
phase, BH mass relations, and BH kicks. GW detections of binaries with
NSs will probe lower-mass stars and NS kicks and the
supernova mechanism. For dynamical formation, we can also probe
cluster properties and their dependence on redshift.

In the past, binary pulsars, supernovae, and gamma-ray burst
observations have been used as constraints on DCO models
\citep[e.g.][]{1998AA...332..173P,OShaughnessy:2005}.  More recently,
studies have explored quantitative, statistical methods for deriving 
constraints and examined the minimum sample sizes needed for
distinguishing between a small set of different isolated-binary
models~\citep{BulikBelczynski:2003,Mandel:2009nx,2010ApJ...725L..91K,2013PhRvD..88h4061O,2013NJPh...15e3027M,Stevenson:2015,2015MNRAS.450L..85M,2015arXiv151004615B}. We
note that the majority of these studies conclude that sample sizes of
order $\sim$100 events are needed for strong constraints.

Before comprehensive quantitative constraints on models become
possible, one might consider whether measurements for this one event
or just a handful of sources would allow us to distinguish between the
two main formation paths: isolated binaries and dynamical
processes. The masses of the BHs in BBH systems from both isolated
binary formation and from clusters depend on the mass spectrum of
single BHs, and thus in both formation channels a range of masses is
expected. For example, the \citet{2015arXiv151004615B} isolated binary
models find detectable BBHs with total masses between 15--20\,\Msun
and $\sim$100\,\Msun (chirp masses up to $\sim$50\,\Msun), the
\citet{2015PhRvL.115e1101R,PhysRevLett.116.029901} cluster models find
chirp masses of 10--22\,\Msun (that could be higher for weaker stellar
winds), and the \citet{2014MNRAS.442.2963K} PopIII BBH mergers have
higher chirp masses (most above 20\,\Msun). The strong dependence on
chirp mass of the distance to which sources can be detected (see
Fig.~\ref{fig:DL_M}) strongly enhances the probability of detecting
these massive BBHs compared to lower-mass objects
\citep{Flanagan:1997sx}.

In view of these predictions, distinguishing between formation in
isolated binaries and through dynamical processes based solely on mass
measurements is unlikely. The situation is similar for mass ratios:
BBH formation through both isolated binary evolution and dynamical
processes tends to favor binaries of roughly comparable masses, within
a factor of
$\sim$\,2~\citep{2015ApJ...806..263D,2015PhRvL.115e1101R,2016arXiv160100007M}.

Initial eccentricities would be very different through the two paths,
but most current predictions are for binaries having circularized by
the time they enter the frequency band of relevance to ground-based
interferometers (see section~\ref{e_spin}). \rev{An accurate
  localization of the source would make it possible to check for the
  presence of nearby clusters. For such localization with GW detectors
  only additional advanced detectors, and a very high signal-to-noise
  \rrev{ratio} would be needed. Alternatively, the discovery of an
  electro-magnetic counterpart could pinpoint the
  position~\citep{GW150914-EMFOLLOW}.  At present,} we are left with two
possibilities, for distinguishing among formation paths: BH spins or
precise determination of the BBH merger rate as a function of
redshift. Detection of spin misalignment would be a strong indication
for dynamical formation, but is challenging, as GW spin measurements
are typically not well constrained
\citep[e.g.][]{GW150914-PARAMESTIM,2014PhRvL.112y1101V,2008CQGra..25r4011V,2008ApJ...688L..61V};
the rates option is challenging too, given the large, overlapping
ranges in the rate predictions from the two paths and their uncertain
redshift evolution. In the future, we may be able to further constrain
models by combining BBH rate constraints with constraints on NS
mergers (even if only upper limits). Consideration of the models
consistent with all these constraints will allow us to make firmer
predictions for detection expectations of other types of EM/GW
binaries involving NSs and white dwarfs.

The BBH population discovered through \event\ also has implications
for other GW detections. \rev{First, before entering the aLIGO/AdV
  band, the BBH systems evolve through the frequency range of
  space-borne GW detectors such as (e)LISA (0.1 -- 10mHz)
  \citep{2013GWN.....6....4A}. Because of the high masses of systems
  like \event, it only takes $\sim$1,000 years to evolve from 2 --
  3~mHz to merger and the systems can be detected not only inside the
  Milky Way, but to distances of $\sim$10~Mpc. These ``heavy''
  stellar-mass BBHs could be plausible (e)LISA sources, if the merger
  rate is at the upper end of the inferred range.}

Second, the expected increase in the merger rate of BBHs
towards higher redshift opens the possibility that the large number of
individually unresolvable high-redshift BBH mergers would instead form a
detectable stochastic background signal.  Such a signal could be probed with aLIGO/AdV detection of, or upper  limits on, the stochastic GW background, as explored
in detail in~\citet{GW150914-STOCHASTIC}.

The possibility that \event\ is produced by a binary of the
first generation PopIII stars may provide a direct link between the
local Universe and the BHs that may have been the seeds that grew into
the supermassive BHs we find in the centers of most galaxies
\citep{2003ApJ...582..559V}. Even if \event\ itself is not a product
of PopIII stars, the confirmation of the high BH masses expected from
the \rev{weaker} stellar winds of low-metallicity stars also supports
the idea that PopIII stars, with even much lower metallicity, may
produce even more massive BHs, unless they become so massive that they
are completely disrupted by pair-instability supernovae (e.g., \citealt{2001ApJ...550..372F,2007Natur.450..390W}).

\section{Conclusions}\label{conclusions}

We have examined the implications of the GW discovery of a BBH
merger in the context of the existing literature on the formation of
BBHs in isolated binaries and in dense stellar environments. Despite
the fact that we have only one firm detection, we can draw several
astrophysical conclusions. 

For the first time we have observational evidence that BBH systems
actually form in nature, with properties such that they merge in the
local Universe. This is a unique confirmation of numerous theoretical
predictions over the past forty years that merging BBHs can form, from
both isolated binaries in galactic fields and from dense stellar
environments. Notably, the measured BH masses in the merging binary
are higher than any of the BH masses dynamically measured reliably
from X-ray binaries. Such ``heavy'' BHs require that they were formed
from massive stars in low-metallicity environments (1/2 \Zsun\ or
lower), given our current understanding of massive-star winds and
their dependence on metallicity. Model rate predictions from both
formation mechanisms are broadly consistent with the BBH merger rate
implied by the \event\ discovery. The relatively extreme models which
either abort the formation of merging BBHs or predict rates lower than
$\simeq$\,1\,\pergpcyr\ are now excluded. Apart from weaker winds at
low metallicities, a significant fraction of BHs must receive low
kicks; survival through common-envelope phases or high rotation in
massive stars may be necessary. We note that the majority of recent
model predictions survive this constraint. Targeted simulations and
additional GW merger detections will be needed to quantify the balance
between BBH formation rate, delay times until merger, and hence BBH
merger rates as a function of redshift. This first BBH discovery
already has implications for a stochastic GW background and for the
potential of observations with a future eLISA-like space mission.

These are the key conclusions we can derive based on the
\event\ properties and the existing DCO astrophysics
literature. Final analysis of this first aLIGO observational run
  may provide additional rate constraints from additional detections
  of BBHs or NS binaries, or in their absence interesting upper limits
  on merger rates of NS binaries. These combined rate constraints
will provide the most stringent quantitative limits on model
predictions. An increased source sample resulting from future GW data
will of course better constrain the merger rates, but will also allow
us to probe the mass distributions and any dependence on redshift. To
go beyond the current, mostly qualitative discussion, and move towards
comprehensive model constraints, it will be important to develop
frameworks that account for observational biases and for appropriate
sampling of the model parameter space including relevant parameter
degeneracies. In closing, we are looking forward to the development of
GW astronomy as a new way of probing the Universe.

\acknowledgments
\section*{Acknowledgements}
The authors gratefully acknowledge the support of the United States
National Science Foundation (NSF) for the construction and operation of the
LIGO Laboratory and Advanced LIGO as well as the Science and Technology Facilities Council (STFC) of the
United Kingdom, the Max-Planck-Society (MPS), and the State of
Niedersachsen/Germany for support of the construction of Advanced LIGO 
and construction and operation of the GEO600 detector. 
Additional support for Advanced LIGO was provided by the Australian Research Council.
The authors gratefully acknowledge the Italian Istituto Nazionale di Fisica Nucleare (INFN),  
the French Centre National de la Recherche Scientifique (CNRS) and
the Foundation for Fundamental Research on Matter supported by the Netherlands Organisation for Scientific Research, 
for the construction and operation of the Virgo detector
and the creation and support  of the EGO consortium. 
The authors also gratefully acknowledge research support from these agencies as well as by 
the Council of Scientific and Industrial Research of India, 
Department of Science and Technology, India,
Science \& Engineering Research Board (SERB), India,
Ministry of Human Resource Development, India,
the Spanish Ministerio de Econom\'ia y Competitividad,
the Conselleria d'Economia i Competitivitat and Conselleria d'Educaci\'o, Cultura i Universitats of the Govern de les Illes Balears,
the National Science Centre of Poland,
the European Union,
the Royal Society, 
the Scottish Funding Council, 
the Scottish Universities Physics Alliance, 
the Lyon Institute of Origins (LIO),
the National Research Foundation of Korea,
Industry Canada and the Province of Ontario through the Ministry of Economic Development and Innovation, 
the National Science and Engineering Research Council Canada,
the Brazilian Ministry of Science, Technology, and Innovation,
the Leverhulme Trust, 
the Research Corporation, 
Ministry of Science and Technology (MOST), Taiwan
and
the Kavli Foundation.
The authors gratefully acknowledge the support of the NSF, STFC, MPS, INFN, CNRS and the
State of Niedersachsen/Germany for provision of computational resources.

\section*{Appendix}\label{appendix}

The sensitivity of the detector network to GW emission from equal-mass
BBHs with non-spinning components is calculated using the following
procedure.  We use a single-detector signal-to-noise ratio threshold
of 8 as a proxy for the detectability of binary mergers by a detector
network; this is a commonly used proxy \citep[e.g.,][]{ratesdoc} which
has been demonstrated to be accurate to within $\sim 10\%$ for
computing surveyed volumes~\citep{GW150914-RATES}.  The curves labeled 2010,
2015--2016, 2017--2018 and 2019+ are computed using, respectively, the
measured noise power spectral density (PSD) of H1 during the S6
science run, the measured noise PSD of H1 during the 2015 science run,
low-end predictions for LIGO noise PSD for the late stages of detector
commissioning, and for design sensitivity runs in the zero detuning,
high power configuration \citep{Aasi:2013wya}.  We use
inspiral-merger-ringdown effective one-body waveforms calibrated to
numerical relativity for these calculations
\citep{Taracchini:2013rva}. The actual sensitivity will depend on the
exact network configuration, the data quality, and the signal
parameters, so the curves in Figure~\ref{fig:DL_M} should be viewed
only as approximations. \rev{In particular, the signal strength and
  detectability generally depend on BH spins.}

The left panel shows the horizon \rev{distance, which is the
  luminosity distance } at which GWs from a face-on, \rev{equal-mass},
overhead binary with the given source-frame total mass (bottom axis)
or chirp mass (top axis) would be detected at a signal-to-noise ratio
of 8; the corresponding redshift is shown on the right vertical axis.
The right panel shows the surveyed detection-weighted comoving volume
$\overline{V}_c$
\begin{equation}
\label{DWSCV}
\overline{V}_c = \int_0^\infty \frac{dV_c}{dz} f_d(z) \frac{1}{1+z} dz\, ,
\end{equation}
where $\frac{dV_c}{dz}$ is computed using the
\citet{2015arXiv150201589P} cosmology and the last factor corrects for
the difference in source and observer clocks. \rev{Because the GW
  strength of signals depends (to within factors $\sim$2) on inclination and
  the detector response depends strongly on sky position,}
$f_d(z)$ is the probability that a binary with the given source-frame
masses at redshift $z$ is louder than the signal-to-noise ratio
threshold of 8 (integrated over isotropic sky locations and
orientations). With this definition, and assuming a constant
volumetric merger rate $\mathcal{R}$ per unit comoving volume per unit
source time, the expected number of detections during an observing run
of duration $T$ is given by $\mathcal{R} \overline{V}_c T$.

\bibliography{AstroBib,cbc-group,macros/GW150914_refs}{}
\bibliographystyle{apj}

\end{document}